\documentclass[journal]{IEEEtran}

\ifCLASSINFOpdf
\else
\fi
\usepackage{hyperref}
\usepackage{array}
\usepackage{mathtools}
\usepackage{amsmath, array}

\usepackage{makecell,multirow,diagbox} 
\usepackage{verbatim}
\usepackage{enumerate}
\usepackage{booktabs}
\usepackage{subfigure}
\usepackage{graphicx}
\usepackage{slashbox,multirow}
\usepackage{epstopdf}
\usepackage{placeins}
\usepackage{algorithmicx}
\usepackage{amssymb}
\usepackage{mathtools}
\usepackage[boxruled]{algorithm2e}
\usepackage{color, colortbl}
\usepackage{siunitx}
\usepackage{cuted}
\usepackage{threeparttable}

\usepackage[noadjust]{cite}

\definecolor{Gray}{gray}{0.85}

\begin{document}

\title{An Effective Methodology for Short-Circuit Calculation of Power Systems Dominated by Power Electronics Converters Considering Unbalanced Voltage Conditions and Converter Limits} 

\author{Jie Song,
Marc~Cheah-Mane,~\IEEEmembership{Member,~IEEE,}
Eduardo~Prieto-Araujo,~\IEEEmembership{Senior Member,~IEEE,}
and~Oriol~Gomis-Bellmunt,~\IEEEmembership{Fellow,~IEEE}


}
\maketitle

\begin{abstract}
 
This paper deals with the challenge of short-circuit calculation for power systems dominated by power electronics converters. A novel methodology has been presented to identify short-circuit equilibrium point of the studied system considering the operation and limitations of power converters. The studied system has been modeled with an element-based steady-state formulation. In particular, the governing equations implemented on converter control, which involve a specific control mode and various potential current-saturation states, are included. Then, an effective and efficient approach has been proposed to identify the short-circuit equilibrium points that satisfies converters limitations. Numerical case studies with VSCs show that the proposed methodology can identify the short-circuit equilibrium point efficiently and accurately with different types and depths of short-circuit fault. The short-circuit calculation results have been validated with dynamic simulations.    
 
\end{abstract}

\begin{IEEEkeywords}
Short-Circuit Calculation, Voltage Source Converter, Current-Saturation, Asymmetrical Fault.
\end{IEEEkeywords}

\section{Introduction}

Modern power systems are increasingly penetrated with power electronics converters as they have been widely adopted in grid integration of renewable generation units and batteries as well as in the applications of high-voltage direct current (HVDC) transmission and flexible AC transmission system (FACTS) \cite{9520270, 7750608, 8302935, 4371538, 5466243, 8986622}. As a consequence, such high penetration of power electronics significantly changes the steady-state characteristics of power systems. This is because compared to conventional loads and generators, power electronics converter could operate in different control modes that introduce non-linear characteristics to the grid . In addition, the converter saturates current in case of being overloaded to protect the semiconductor device. Such current-saturation modifies the converter operation and should be considered in the steady-state analysis of the studied system \cite{Strezoski2019}. This is especially critical for short-circuit calculation as several or all converters are likely to operate with a saturated current. 

Short-circuit calculation is a fundamental but essential task in power system computational analysis. Short-circuit equilibrium points identified in different fault scenarios provide the basis for secure sizing of electrical installations and proper tuning of protection schemes \cite{c37.9, IEC60826, Jones2012}. The grid-support control of power electronics converters is also designed based on the short-circuit calculation results \cite{Islam2020}. In addition, the identified short-circuit equilibrium point can be adopted as initial conditions for EMT and small-signal analyses related to fault conditions \cite{8477141,MAHSEREDJIAN20071514}. 

Conventional programs for short-circuit calculation usually adopt the Thévenin or Norton equivalent to characterize the studied network from the fault location \cite{Das2007}. However, these linearized grid-equivalents disregards the complex characteristics of converter control and are not suitable for the system with high penetration of power electronics \cite{STD3002,SC_WhitePaper,5589677}. The IEC 60909 standard suggests to model the power converters as current sources for short-circuit calculation \cite{IEC2016}. However, it has not been clearly stated how to obtain the current angle for a specific fault condition \cite{Aljarrah2019, 9634593}.    

The short-circuit analysis have been widely reported for power systems dominated by power converters in the literature. A fault ride through scheme has been proposed in \cite{6045345} for wind farms with type IV wind turbines. A fault detection and localization algorithm has been presented for distribution system dominated by power converters in \cite{7080908}. Short-circuit current contributions from converter-interfaced generation units have been studied in \cite{6629360,Neumann2012a,7131563}. The studies presented in the mentioned references \cite{6045345,7080908,6629360,Neumann2012a,7131563} are based on dynamic (EMT) simulations for each short-circuit equilibrium point to be analyzed. However, a steady-state approach is more efficient for short-circuit calculation.       

The existing literature contains limited solutions for short-circuit calculation considering the operation and limitations of power electronics converters. A load-flow based approach is proposed to calculated the fault current contribution from power converters in \cite{5589981, 8881660}. A sequence-domain phasor model has been proposed in \cite{Kauffmann2019, PSRCC, 8013760} for short circuit calculation of VSC dominated systems where the grid-support control and current limitation are considered. A grid equivalent has been presented in \cite{Furlaneto2021a} where the power converters are represented with a voltage controlled current source for short-circuit calculation. The methodologies presented in the references consider only the converter operation in PQ control. A systematic approach that covers different converter control modes (\textit{e.g.} PV or grid-forming control) is still needed.     

The authors of this paper have developed a different approach for short-circuit calculation of power systems penetrated with power electronics with a balanced voltage condition \cite{9420721,9504581,9770768,SONG2022108352}. The studied system is modeled with an element-based formulation where the converters equivalent model is included. Then, the short-circuit equilibrium points are identified by solving the established systems of equations corresponding to different combinations of converters' current-saturation states. Based on that, a further step is to extend the proposed formulation and methodology to an unbalanced voltage condition in order to carry out short-circuit calculation with different types of fault.   

This paper presents a methodology for effective and efficient short-circuit calculation of power systems with penetration of power electronics where the operation and limits of power converters are considered. The presented methodology has been tested with several VSC-based numerical case studies. The short-circuit calculation results have been validated with dynamic simulations. In this paper, the studied power systems are presented with a sequence expression and using per-unit values. 

\section{Steady-State Formulation for Short-Circuit Calculation}

\subsection{VSC Equivalent Model}\label{sec:VSC modeling}

\begin{figure*}[!htb]
	\centering
	\includegraphics[width=0.7\textwidth]{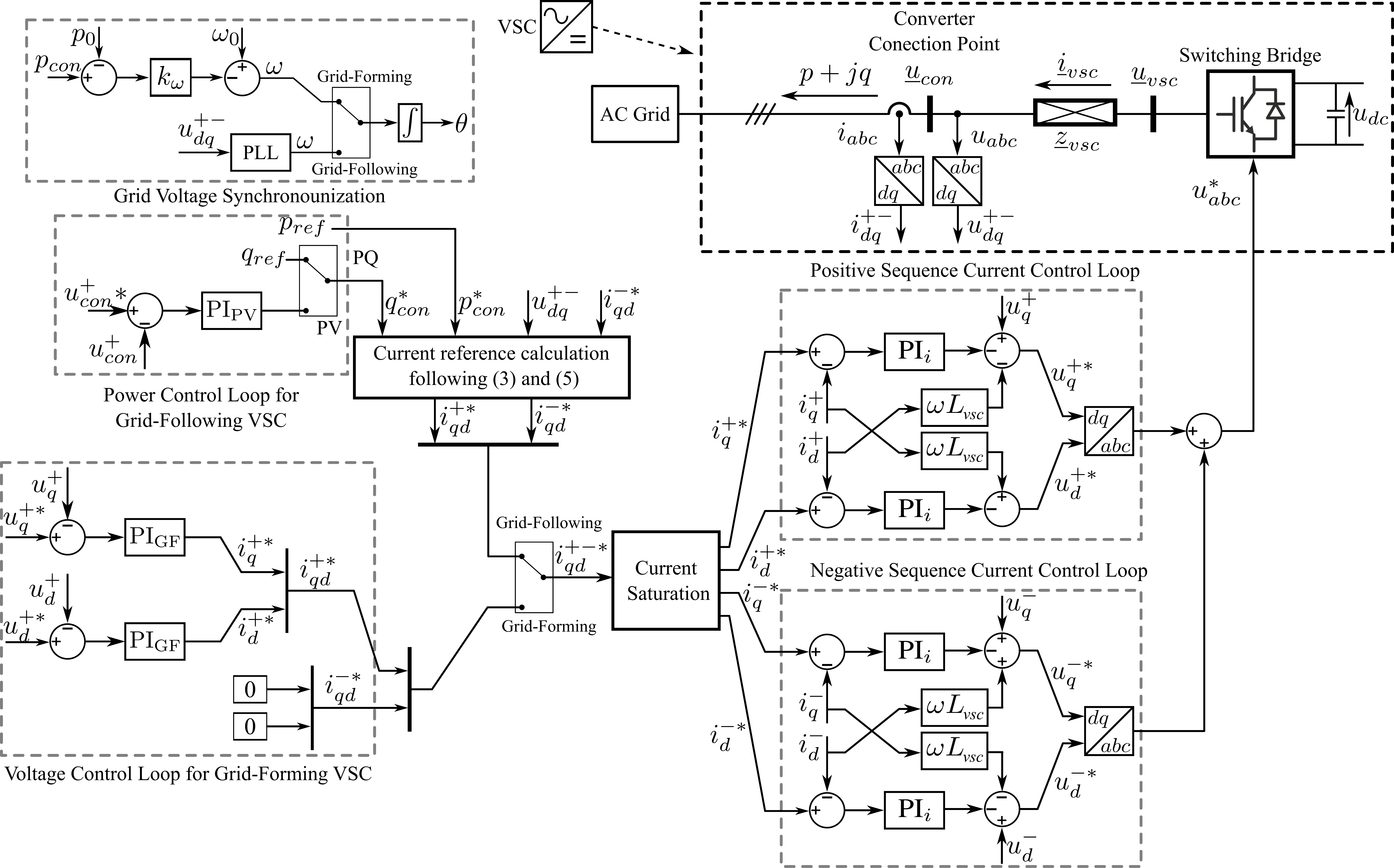}
	\caption{VSC equivalent diagram and the control loop scheme}
	\label{fig:VSC equivalent diagram}
\end{figure*}

The VSC is expressed with a voltage source and a series reactor for steady-state analysis which regulates the power exchange with the AC grid at the Converter Connection Point as shown in Fig.~\ref{fig:VSC equivalent diagram}. The VSC operation is modeled with a sequence expression, which can be converted from or to natural frame three-phase components following the Fortescue transformation shown in the Appendix. 

The VSC equivalent model is presented with a three-phase configuration in this paper as such configuration is typically adopted in power engineering applications \cite{5765677, 5443484}. However, the rest part of the power system can be analyzed in either three- or four-wire configurations. Therefore, the zero-sequence current can not be injected by the VSC. Also, the zero-sequence voltage does not contribute to the converter power exchange with the AC grid. As the consequence, the VSC active power injection can be expressed as follows:
\begin{equation}
	\label{eq:active power injection_sequential}
	p=\underbrace{p_{con}}_{\bar p}+\underbrace{\cos(2\omega t+2\theta_0)p_{cos}+\sin(2\omega t+2\theta_0)p_{sin}}_{\tilde p}
\end{equation}

\noindent where $\omega$ is the angular frequency of the studied system, $\theta_0$ is the initial phase angle for the angular reference, $\bar p$ and $\tilde p$ respectively denominate the constant and oscillating elements of the active power, and the three active power components can be expressed as \cite{6017130, Wang2011, Xu2005}: 
\begin{equation}
	\label{eq:active power_sequential elements}
	\begin{cases}
		p_{con}=\underbrace{{u_x^+ i_x^+} + {u_y^+ i_y^+}}_{p_{con}^+} + \underbrace{{u_x^- i_x^-} + {u_y^- i_y^-}}_{p_{con}^-}\\[2pt]
		p_{cos}={u_x^+ i_x^-} + {u_y^+ i_y^-} + {u_x^- i_x^+} + {u_y^- i_y^+}\\[2pt]
		p_{sin}=-{u_x^+ i_y^-} + {u_y^+ i_x^-} + {u_x^- i_y^+} - {u_y^- i_x^+}
	\end{cases}
\end{equation}

\noindent where $u$ is the CCP voltage, $i$ is the converter current injection, superscripts $+$ and $-$ respectively denote the positive and negative sequence components, $p_{con}^+$ and $p_{con}^-$ are the constant active power elements respectively corresponding to the positive and negative sequence current. 

Similarly, the VSC reactive power injection can be expressed as follows \cite{6017130, Wang2011, Xu2005}: 
\begin{equation}
	\label{eq:reactive power injection_sequential}
	q=\underbrace{q_{con}}_{\bar q}+\underbrace{\cos(2\omega t+2\theta_0)q_{cos}+\sin(2\omega t+2\theta_0)q_{sin}}_{\tilde q}
\end{equation}

\noindent where the constant and oscillating reactive power elements can be expressed as:
\begin{equation}
	\label{eq:reactive power_sequential elements}
	\begin{cases}
		q_{con}=\underbrace{{u_y^+ i_x^+} - {u_x^+ i_y^+}}_{q_{con}^+}  \underbrace{ + {u_y^- i_x^-} - {u_x^- i_y^-}}_{q_{con}^-}\\[2pt]
		q_{cos}={u_x^+ i_y^-} - {u_y^+ i_x^-} + {u_x^- i_y^+} - {u_y^- i_x^+}\\[2pt]
		q_{sin}={u_x^+ i_x^-} + {u_y^+ i_y^-} - {u_x^- i_x^-} - {u_y^- i_y^+}
	\end{cases}
\end{equation}

\noindent where $p_{con}^+$ and $p_{con}^-$ are the constant reactive power elements respectively corresponding to the positive and negative sequence current. 

Different control strategies can be adopted by the VSC. In this paper, the VSC operation in grid-following, which includes PQ and PV modes, and grid-forming control are presented to exemplify the VSC equivalent model. Also, the current-saturation scheme is employed in the converter control as shown in Fig.~\ref{fig:VSC equivalent diagram}. Such current-saturation modifies the converter operation in case of being overloaded, which results in a piece-wise format of converter equivalent model. 

\subsubsection{PQ control}  

The VSC in PQ control manipulates the positive sequence current to regulate the constant elements in active and reactive power injections, $p_{con}$ and $q_{con}$, following the reference value, while the oscillating power elements, $\tilde p$ and $\tilde q$, are not being controlled directly \cite{6017130}. The negative sequence current, $\underline{i}^-$, is set to zero in order to impose a symmetrical current injection even during unbalanced faults \cite{8894890, PSRCC, Jia2018a, 6520231, 4558266, 7102165}. 

For the converter in PQ control, both the active and reactive current reference in positive sequence can be achieved in normal operation (unsaturated state, USS). When the current hits the limitation, the priority is given to reactive current while the active element is reduced (partially-saturated state, PSS) . When the fault is deep, both active and reactive current references cannot be achieved due to the converter current limitation. In such case, the converter prioritize active or reactive current element with full capacity in positive sequence (fully-saturated state, FSS). Therefore, the VSC operation with PQ control can be divided into three current-saturated states (USS, PSS and FSS), which are expressed as follows \cite{9420721,9504581,9770768, SONG2022108352}:
\begin{equation}
	\label{eq:VSC_PQ}
	\begin{cases}
		p_{con}=p_{ref};\;q_{con}=q_{ref};\;\underline{i}^-=0&\rm{if}\;\rm{USS}\\
		q_{con}=q_{ref};\;i^+=i_{vsc}^{\max};\;\underline{i}^-=0&\rm{if}\;\rm{PSS}\\
		p_{con}=0;\;i^+=i_{vsc}^{\max};\;\underline{i}^-=0&\rm{if}\;\rm{FSS}
	\end{cases}
\end{equation}

\noindent where $p_{ref}$ and $q_{ref}$ are reference value for active and reactive power (the constant elements), $i_{vsc}^{\max}$ is the VSC nominal current. The reactive current element is prioritized in current-saturated states (PSS or FSS). However, the active current can be prioritized with a similar formulation. 

The VSC tracks constant dispatched power references, $p_{disp}+jq_{disp}$, in normal operation. When the short-circuit fault happens, assuming the fault can be detected instantaneously, the reactive power controller output, $i_{d0}$, will be frozen and a voltage-droop grid support value will be added following the grid codes \cite{PSRCC, Hagh2019, Islam2020}. The frozen value, $i_{d0}$, is depending on the pre-fault operation point of the studied system and is considered as a given constant in this paper for short-circuit calculation. The active power reference $p_ref$ remains unchanged in different conditions. Then, the power references for the VSC in PQ control can be expressed as follows \cite{9504581,9770768, SONG2022108352}: 
\begin{equation}
	\footnotesize
	\label{eq:VSC_PQ power reference}
	\begin{cases}
		p_{ref}=p_{disp};\;q_{ref}=q_{disp}&\rm{if}\;\rm{fault}=0\\
		p_{ref}=p_{disp};\;q_{ref}=u^+[i_{d0}+k_{isp}(u_{ref-gs}-u^+)]&\rm{if}\;\rm{fault}=1
	\end{cases}
\end{equation}

\noindent where $k_{isp}$ is the voltage-droop gain of grid-support current and $u_{ref-gs}$ is the positive sequence voltage reference for grid-support control during the fault. 

\subsubsection{PV Control} 
The VSC in PV control also tracks a dispatched value for the constant element of active power, $p_{con}$, while the constant element of reactive power element, $q_{con}$, is adjusted to regulate the positive sequence voltage magnitude, $u^+$. The three current-saturation states (USS, PSS and FSS) also apply to the converter in PV control, which results in the equivalent model as follows \cite{9420721,9504581}:  
\begin{equation}
	\label{eq:VSC_PV}
	\begin{cases}
		p_{con}=p_{disp};\;u^+=u_{ref-PV};\;\underline{i}^-=0&\rm{if}\;\rm{USS}\\
		u^+=u_{ref-PV};\;i^+=i_{vsc}^{\max};\;\underline{i}^-=0&\rm{if}\;\rm{PSS}\\
		p_{con}=0;\;i^+=i_{vsc}^{\max};\;\underline{i}^-=0&\rm{if}\;\rm{FSS}
	\end{cases}
\end{equation}

\noindent where $u_{ref-PV}$ is the positive-sequence voltage magnitude reference for the VSC in PV control. The reactive current element in positive sequence (which is corresponding to voltage control) is prioritized in a current-saturated state over the active current, which is the same as the converter in PQ control as expressed in \eqref{eq:VSC_PQ}.

\subsubsection{Grid-Forming} 

The grid-forming converter regulates the CCP positive sequence voltage and the studied system frequency in USS. When the current limit is reached, the grid-forming converter inject a positive sequence current with full capacity while the negative sequence current is always set to zero in order to impose a symmetrical current \cite{RMGF}. For the system with multiple grid-forming units, a $p-\omega$ droop control is adopted to implement the distributed slack for power sharing purpose \cite{Zhang2021, Rokrok2020}. The partially-saturated state (PSS) does not apply to a grid-forming converter as the active or reactive power elements are not being controlled directly. Therefore, the grid-forming VSC operation can be divided into two current-saturation states (USS and FSS) as follows \cite{9504581}:
\begin{equation}
	\label{eq:VSC_GF}
	\begin{cases}
		{u}^+={u}_{ref-GF};\;\underline{i}^-=0;\omega=\omega_0-k_{\omega}(p_{con}-p_{0})&\rm{if}\;\rm{USS}\\
		{i}^+=i_{vsc}^{\max};\;\underline{i}^-=0;\omega=\omega_0-k_{\omega}(p_{con}-p_{0})&\rm{if}\;\rm{FSS}
	\end{cases}
\end{equation}

\noindent where ${u}_{ref-GF}$ is the positive sequence voltage magnitude reference for the grid-forming converter, $\omega_0$ is the nominal grid  frequency, $k_{\omega}$ and $p_0$ are the droop gain and dispatched active power for frequency droop control. The frequency droop gain $k_{\omega}$ can be set to zero for micro-grid with only one grid-forming converter to impose a constant frequency, \textit{i.e.} $\omega=\omega_0$.  

It should be noticed that various options can be adopted for VSC control in different applications. For example, the active current elements can be prioritized for a grid-following converter in a current-saturated state \cite{9692467}; the converter in PQ control can regulate the oscillating power elements \cite{4153398}; negative sequence current can be injected by the converter if it is required by the grid code \cite{6672880}. These different options are not explicitly expressed in this paper due to the limited space. However, the system formulation and the methodology for short-circuit calculation presented in this paper can be adapted to handle these different options of VSC control as long as the corresponding converter equivalent model is given. 

\subsection{Modeling of Power Systems with penetration of power electronics}\label{sec:power system formulation}

The passive grid components (\textit{e.g.} transmission lines, transformers, load impedance, short-circuit fault impedance) of the studied system are modeled with the admittance matrix in a sequence format as follows: 
\begin{equation}
	\label{eq:circuits admittance matrix}
	\scriptsize
	\underbrace{\left[ {\begin{array}{*{20}{c}}
				{{\underline{i}_1^+}}\\
				\vdots \\
				{{\underline{i}_D^+}}\\[5pt]
				{{\underline{i}_1^-}}\\
				\vdots \\
				{{\underline{i}_D^-}}\\[5pt]
				{{\underline{i}_1^0}}\\
				\vdots \\
				{{\underline{i}_D^0}}\\
		\end{array}} \right]}_{\mathbf{I}} = \underbrace{\left[ {\begin{array}{*{20}{c|c|c}}
				\mathbf{Y^{++}}& \mathbf{Y^{+-}} & \mathbf{Y^{+0}}\\
				\cmidrule[0.4pt]{1-3}
				\mathbf{Y^{-+}} & \mathbf{Y^{--}} & \mathbf{Y^{-0}}\\
				\cmidrule[0.4pt]{1-3}
				\mathbf{Y^{0+}} & \mathbf{Y^{0-}} & \mathbf{Y^{00}}\\
		\end{array}} \right]}_{\mathbf{Y}}
	\underbrace{\left[ {\begin{array}{*{20}{c}}
				{{\underline{u}_1^+}}\\
				\vdots \\
				{{\underline{u}_D^+}}\\[5pt]
				{{\underline{u}_1^-}}\\
				\vdots \\
				{{\underline{u}_D^-}}\\[5pt]
				{{\underline{u}_1^0}}\\
				\vdots \\
				{{\underline{u}_D^0}}\\
		\end{array}} \right]}_{\mathbf{U}}
\end{equation}

\noindent where $D$ is the number of total buses in the studied grid, $\underline{u}_d^{+-0}$ is the voltage at bus $d$ for $d \in [1,D]$. The admittance matrix, $\mathbf{Y}$, has the dimension of $3D \times 3D$. 

An additional set of constraint equations, $H$, are employed to model the operation of power electronics and non-power electronics elements. In particular, power electronics can operate in various current-saturation state, which results in different equations in their equivalent model as expressed in Section~\ref{sec:VSC modeling}. Therefore, power electronics converters are modeled with different equations corresponding to different current-saturation states. In particular, combinations number of all possible converters' current-saturation states can be expressed as: $F = \prod\nolimits_{m = 1}^{M}{x_m}$ where $x_m$ is the number of of current-saturation states of converter $m$ \cite{9420721,9504581}. In particular, $x_m$ is equal to 3 for a grid-following converter and 2 for a grid-forming converter following the equivalent model expressed in this paper. Then, each subset of equations of $H$, $H_f$, can be obtained to define the operation of all power electronics and non-power electronics elements in the studied system corresponding to each combination $f \in [1,F]$ such that \cite{9420721,9504581,9770768, SONG2022108352}:
\begin{equation}
	\label{eq:Hf constraints}
	H_f={\left[ {\begin{array}{*{20}{c}}
				{{h_{1-f}}}& \cdots & {{h_{m-f}}} & \cdots &{{h_{M-f}}}
		\end{array}} \right]^T}
\end{equation}
\noindent where $M$ denotes the number of all elements (both power electronics and non-power electronics) in the studied system and subset of equations $h_{m-f}$ models the operation of element $m$ corresponding to combination $f$ for $m \in [1,M]$.   

Non-power electronics elements always operate in the same state without being saturated. Therefore, their equivalent model will not be modified for different combinations $f$ (\textit{i.e.} number of states $x_m=1$). The equations modeling the most typical non-power electronics elements are expressed as follows: 
\begin{equation}
	\footnotesize
	\label{eq:non-PE current injection}
	\begin{cases}
		h_{m-f}\coloneqq\left[ {\begin{array}{*{20}{c}}
				{\underline{i}_{d,m}^{+}}= \frac{{{\underline{u}_{th,m}^{+}}-{\underline{u}_{d}^{+}}}}{{\underline{z}_{th,m}}}\\
				{\underline{i}_{d,m}^{-}}= \frac{{-{\underline{u}_{d}^{+}}}}{{\underline{z}_{th,m}}}\\
				{\underline{i}_{d,m}^{0}}= \frac{{-{\underline{u}_{d}^{+}}}}{{\underline{z}_{th,m}}}
		\end{array}} \right]&{\rm{if}}\;{\rm{Th\acute{e}venin}}\;{\rm{equivalent}}\\[12pt]
			h_{m-f}\coloneqq\left[ {\begin{array}{*{20}{c}}
				{\underline{u}_{d,m}^{+}}= {\underline{u}_{ref,m}^{+}}\\
				{\underline{u}_{d,m}^{-}}= 0\\
				{\underline{u}_{d,m}^{0}}= 0
		\end{array}} \right]&{\rm{if}}\;{\rm{Slack}}\;{\rm{bus}}\;({\rm{non}}\;{\rm{PE}})\\[10pt]
		h_{m-f}\coloneqq{\underline{i}_{d,m}^{ph}}=\left(\frac{p_{ref,m}+jq_{ref,m}}{{\underline{u}_{d}^{ph}}}\right)^* &{\rm{if}}\;{\rm{PQ}}\;{\rm{node}}\;({\rm{non}}\;{\rm{PE}})\\[12pt]
		h_{m-f}\coloneqq\left[ {\begin{array}{*{20}{c}}
				{u}_{d}^{+}=u_{ref,m}\\
				{\rm{Re}}\left({\underline{u}_{d}}{\underline{i}_{d,m}^{ph*}}\right)=p_{ref,m}
		\end{array}} \right]&{\rm{if\;PV\;node}}\;({\rm{non}}\;{\rm{PE}})
	\end{cases}
\end{equation}  

\noindent where subscript $ph \in \{a,b,c\}$ denotes the element for each phase in a natural three-phase frame.

In the presented examples, it is assumed the Thévenin equivalent voltage, $\underline{u}_{th,m}$, and the slack bus voltage contain only positive sequence component, the PQ node follows the same power references for each phase, the PV node tracks the active power references for each phase and imposes a symmetrical voltage following the reference value \cite{1490591}. It should be noticed that non-power electronics elements can be also modeled differently considering various options. These different options are not explicitly expressed in this paper but they can be equally covered by the proposed system formulation and methodology for short-circuit calculation. 

Then, the system of equations, $SE_f$, can be defined to model the studied power system corresponding to each combination $f \in [1,F]$ short-calculation such that: 
\begin{equation}
	\small
	\label{eq:system of equations}
	SE_f \coloneqq \begin{cases}
		\mathbf{I}=\mathbf{Y}\mathbf{U}\\[2pt]
		H_f={\left[ {\begin{array}{*{20}{c}}
					{{h_{1-f}}}& \cdots & {{h_{m-f}}} & \cdots &{{h_{M-f}}}
			\end{array}} \right]^T}
	\end{cases}
\end{equation}

\noindent where the sequence admittance matrix, $\mathbf{Y}$, is the same for all combinations while the corresponding set of equations $H_f$ will be selected to model operation of all elements in the studied system for each combination $f$.

\section{Methodology to Identify Short-Circuit Equilibrium Point}\label{sec:methodology}

This Section introduces an effective and efficient methodology to identify the short-circuit equilibrium point that satisfies all converters' operation limits. The proposed methodology contains two levels: an outer loop which updates the current-saturation states of all power converters iteratively and an inner loop which defines and solves the set of equations modeling the studied system following the combination number $f$ given by the outer loop. 

\subsection{Iteration of Converters' Current-Saturation States} 

Power converters' current-saturation states are usually uncertain with a specific short-circuit fault scenario \cite{Kauffmann2019, PSRCC, 9420721,9504581}. In other words, the short-circuit equilibrium point might be obtained from any of the $F$ possible combinations. A possible solution approach is to go through all the $F$ combinations and solve the corresponding systems of equations, $SE_{f} \; \forall f\in[1,F]$, in order to obtain the short-circuit equilibrium point that satisfies converters' limits \cite{9420721,9504581}. However, this approach is not efficient for the system with several converters as it takes long computing time to traverse all the $F$ possible combinations. 

As an alternative, the authors of this paper proposed an iterative methodology to identify the current-saturation states of power converters during the fault and obtain the short-circuit equilibrium point, which is suitable for converter in PQ control and limited with a balanced fault condition \cite{9770768,SONG2022108352}. In this paper, this iterative methodology has been extended to include converters operation in various control modes (PQ, PV and grid-forming) and adapted to the potential unbalanced voltage conditions. 

The proposed methodology tested only selected combinations of converters' current-saturation states for short-circuit calculation. a system of equations, $SE_f$, is defined and solved for each iteration, $n_{t} \in [1,N_t^{\max}]$, where $N_t^{\max}<F$ is the maximum number of iterations allowed. The combinations for the upcoming iteration, $n_t+1$, is obtained based on the solution of $SE_f$, $sol$, from the current iteration, $n_t$. In particular, only the current-saturation states of those converters whose operation limits are violated will be modified in the upcoming iteration. 

\begin{figure}[!htb]
	\centering
	\includegraphics[width=0.36\textwidth]{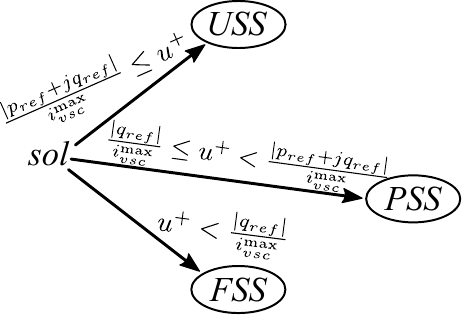}
	\caption{Update Current-Saturation States-PQ}
	\label{fig:Iteration PQ}
\end{figure}

The rule to iteratively update the converters' current-saturation states varies from different converter control modes, which is defined based on converter's operation limits for each control mode and various current-saturation states. In particular, the current-saturation states of converters in PQ control are updated based on the limits in terms of the positive sequence voltage magnitudes at the CCP, which is summarized as shown in Fig.~\ref{fig:Iteration PQ}. In particular, the converter in PQ control can operate in USS for a voltage close to the nominal value as the power references, $p_{ref}+jq_{ref}$, can be achieved without violating the current limit. However, since the grid voltage is commonly reduced during the short-circuit fault, the VSC is likely to operate in PSS or FSS with a limited current \cite{9770768,SONG2022108352}. 

\begin{figure}[!htb]
	\centering
	\includegraphics[width=0.46\textwidth]{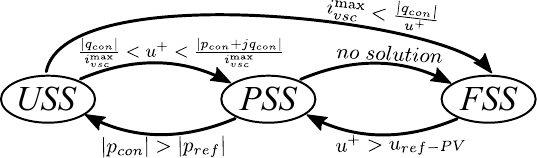}
	\caption{Update Current-Saturation States-PV}
	\label{fig:Iteration PV}
\end{figure}

The iteration rule for the converter in PV mode associates both the obtained solution and the current-saturation state tested in the previous iteration, which is summarized as shown in Fig.~\ref{fig:Iteration PV}. In particular, for the solution obtained for USS, the PV converter will be switched to a current saturated state (PSS or FSS) if the nominal current has been exceeded. For the solution obtained corresponding to PSS, the converter will be switched to USS in the next iteration if the injected active power element, $p_{con}$, has a higher magnitude than its reference value, $p_{ref}$. During a severe fault, the established system of equations does not have solution corresponding to PSS operation of a PV converter. This is because a high amount of reactive current injection will be  required in order to maintain the grid voltage, which is not possible under the converter current limitation. In such case, the converter will be identified as fully saturated (FSS) in the upcoming iteration to fulfill the converter limits. Finally, if a solution obtained for FSS gives a positive sequence voltage even higher than the reference value, the converter will be switched back to PSS. 

\begin{figure}[!htb]
	\centering
	\includegraphics[width=0.24\textwidth]{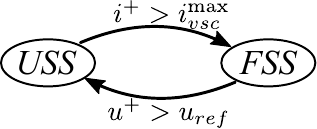}
	\caption{Update Current-Saturation States-Grid Forming}
	\label{fig:Iteration GF}
\end{figure}

Similarly, the adopted iteration rule for the grid-forming converter is shown in Fig.~\ref{fig:Iteration GF}. The operation limits are set based on injected current magnitude for USS and on positive sequence voltage magnitude for FSS. Once the operation limits are exceeded, the current-saturation state of a grid-forming converter will be modified in the upcoming iteration. 

In this paper, a function $DS$ is defined to update converters's current-saturation states applying the iterative rules shown in Figs.~\ref{fig:Iteration PQ},~\ref{fig:Iteration PV}~and~\ref{fig:Iteration GF} such that:
\begin{equation}
	\label{eq:DS function}
	X_{n_t+1}=DS(sol, X_{n_t})	
\end{equation}

\noindent where $X_{n_t}$ is a set current-saturation states of all power converters in the studied system corresponding to an iteration $n_t$ such that:  
\begin{equation}
	\label{eq:iteration state set}
	X_{n_t}=\left[ {\begin{array}{*{20}{c}}
			{{x_{n_t}^1}}& \cdots & {{x_{n_t}^m}} & \cdots &{{x_{n_t}^M}}
	\end{array}} \right]
\end{equation}

\noindent where ${x_{n_t}^m}$ is the current-saturation states of converter $m \in [1,M]$ and corresponding to a iteration number $n_t$. 

\subsection{Overall Process for Short-Circuit Calculation}

The methodology of the overall process for short-circuit calculation is summarized as shown in Algorithm~\ref{algorithm:short-circuit calculation efficient}. The input required by the methodology includes: admittance matrix of the studied system circuit, $\mathbf{Y}$,
constraint equations modeling the power electronics and non-power electronics device operation, $H$, the function to update converters' current-saturation states, $DS$, an initial set of converters' current saturation states, $X_{0}$, and the maximum allowed iteration number, $n_t^{\max}$.

The short-circuit calculation starts with an initial set of converters' current-saturation states, $X_0$. In this paper, such initial set is selected corresponding all converters operated in USS for example. In each iteration, the system of equations, $SE_f$, is defined based on the combination number, $f$, which is obtained from the tested set of converters' current-saturation states, $X_{n_t}$. In this paper, the Levenberg-Marquardt algorithm is adopted to solve the established system of equations, $SE_f$, which is implemented using the $fsolve$ function in MATLAB. However, other iterative solver can be also adopted depending on the convenience of implementation in different applications.    

Based on the obtained solution of the system of equations $SE_f$, $sol$, the current-saturation states of the upcoming iteration will be updated using the predefined function $DS$, which is expressed in \eqref{eq:DS function}. In case all the converters' current-saturation states are not modified in an iteration $n_t$, \textit{i.e} $X_{n_{t}+1}=X_{n_t}$, the obtained solution can be identified as the short-circuit equilibrium point that satisfies all converters' operation limits. Therefore, the calculation program will be terminated and the identified equilibrium point will be returned as the short-circuit calculation result. 

In addition, the tested combination number in each iteration, $f$, will be recorded. The calculation will be interrupted when the updated combination number is repeated with any of the combinations tested previously to avoid the endless loop (except for the case where the updated iteration number is the same as in the current iteration, which indicate the short-circuit equilibrium point has been identified). In such case, a new set of converters' current-saturation state will be assigned in order to continue the calculation program. This new set $X_{new}$ can be defined by the user or randomly selected from the combinations that have not been tested yet \cite{9770768,SONG2022108352}.     

\begin{algorithm}
	{
		\SetKwData{Left}{left}\SetKwData{This}{this}\SetKwData{Up}{up}\SetKwFunction{Union}{Union}\SetKwFunction{FindCompress}{FindCompress}\SetKwInOut{Input}{input}\SetKwInOut{Output}{output}
		\Input{Admittance matrix $\mathbf{Y}$, Set of constraint equations $H$, function $DS$, initial set of converters' saturation states $X_0$, maximum iteration number $n_t^{\max}$}
		\Output{Identified equilibrium (eq.) point $EP$}
		\BlankLine
		\Begin{
			$n_t=0$; $X_1=X_0$\;
			\For{$n_t \leftarrow 1$ \KwTo $n_t^{\max}$}{
				define $f$ with $X_{n_t}$;
				$SE_f \coloneqq [\mathbf{Y}, H_f]$\;
				$sol=fsolve(SE_{f})$; 
				$X_{n_t+1}=DS(sol,X_{n_t})$\;
				$F_t(n_t)=f$; {\small \tcp{save tested comb. No.}}
				\If{$X_{n_t+1}==X_{n_t}$}{
					$EP=sol$; \quad \KwRet{$EP$}\;
					 \textbf{break}; \quad {\small \tcp{Eq. identified}}
				}
			\If{$find(f \in F_t)$}{
				{\small $X_{n_t+1}=X_{new}$; \tcp{avoid endless loop}}
			}
			}
			
				\If{$EP==null$}{
					\KwRet{'No Equilibrium Point'}\;
				}
			
		}
	}
	
	\caption{Identification of Short-Circuit Equilibrium Point\vspace{1ex}}\label{algorithm:short-circuit calculation efficient}
\end{algorithm}

\section{Case Studies}\label{sec:case study}

\subsection{Test System 1}

\begin{figure}[!htb]
	\centering
	\includegraphics[width=0.36\textwidth]{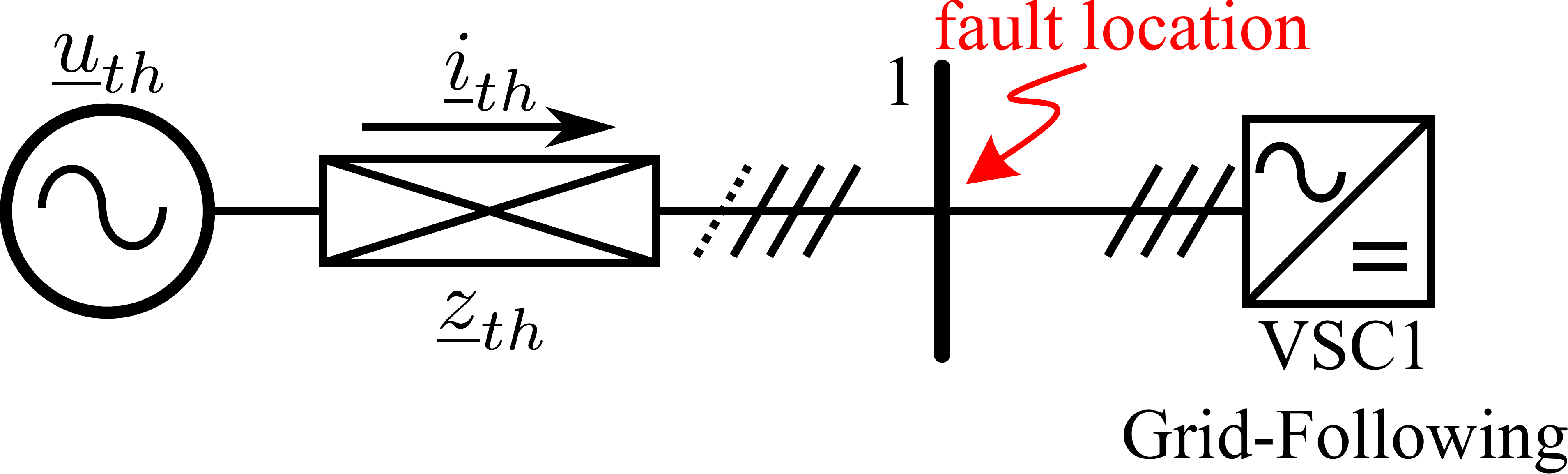}
	\caption{Scheme of Test System 1}
	\label{fig:Test1 Scheme}
\end{figure}

Test System 1 is formed by the AC main grid, which is expressed with the Thévenin equivalent, and a VSC in PQ control with the scheme shown in Fig.~\ref{fig:Test1 Scheme}. The short-circuit fault is tested at the CCP of the converter. Different types of fault (3P2G, P2P and 1P2G)  have been implemented with the fault impedance of $\underline{z}_{ft}=j0.1$ pu. The obtained short-circuit equilibrium points following the methodology shown in Algorithm~\ref{algorithm:short-circuit calculation efficient} are listed in Table~\ref{table:Test1 Results}. 

\begin{table}[!htb]
	\begin{center}
		\caption{Identified Short-Circuit Equilibrium Points of Test System 1}\label{table:Test1 Results}
		\scriptsize
		\begin{threeparttable}
			\begin{tabular}{m{1.6cm}<{\centering}|m{1.7cm}<{\centering}m{1.7cm}<{\centering}m{1.7cm}<{\centering}}
				\toprule
				Type of Fault & 3P2G & P2P & 1P2G   \\
				\midrule
				No. of Ite. $n_t$ & 2 & 2 & 2   \\
				\hline
				$t_{comp}$ ($s$) & 0.02 & 0.02 & 0.02   \\
				\hline
				\scalebox{0.95}{VSC1 State} & FSS & FSS & PSS   \\
				\hline
				$\underline{u}_{1}^+$ & $0.550\angle 2.57\si{\degree}$ & $0.733\angle 0.5\si{\degree}$ & $0.896\angle 4.3\si{\degree}$  \\
				\hline
				$\underline{u}_{1}^-$ & $0$ &\scalebox{0.95}{$0.368\angle -122.4\si{\degree}$} & \scalebox{0.95}{$0.180\angle -179.2\si{\degree}$}  \\
				\hline
				$\underline{u}_{1}^0$ & $0$ & $0$ & \scalebox{0.95}{$0.180\angle -179.2\si{\degree}$}  \\
				\hline
				$\underline{u}_{1}^a$ & $0.317\angle 2.6\si{\degree}$ & $0.356\angle -29.6\si{\degree}$ & $0.310\angle 6.5\si{\degree}$  \\
				\hline
				$\underline{u}_{1}^b$ & \scalebox{0.95}{$0.317\angle -117.4\si{\degree}$} &\scalebox{0.95}{$0.377\angle -89.4\si{\degree}$} & $0.621\angle 63.7\si{\degree}$  \\
				\hline
				$\underline{u}_{1}^c$ & $0.317\angle 112.6\si{\degree}$ & $0.635\angle 119.5\si{\degree}$ & $0.621\angle -56.3\si{\degree}$  \\
				\hline
				\scalebox{0.95}{$p_{con1}+jq_{con1}$} & $0+j0.550$ & $0+j0.733$ & $0.632+j0.635$  \\
				\bottomrule
			\end{tabular}
		\end{threeparttable}
	\end{center}
\end{table}

The short-circuit equilibrium points are identified in the second iteration in this case for different types of fault with the total computing time of 0.02 $s$. The VSC is operating in FSS for 3P2G and P2P faults. During the 1P2G fault, the converter is identified in PSS. Also, oscillating elements appear in the VSC1 power injections during the unbalanced fault (1P2G and P2P) as the converter only regulates the constant power elements in this case as expressed in \eqref{eq:VSC_PQ}. 

\begin{figure}[!htb]
	\centering
	\subfigure[Fault voltage-3P2G.]{
		\includegraphics[width=0.225\textwidth]{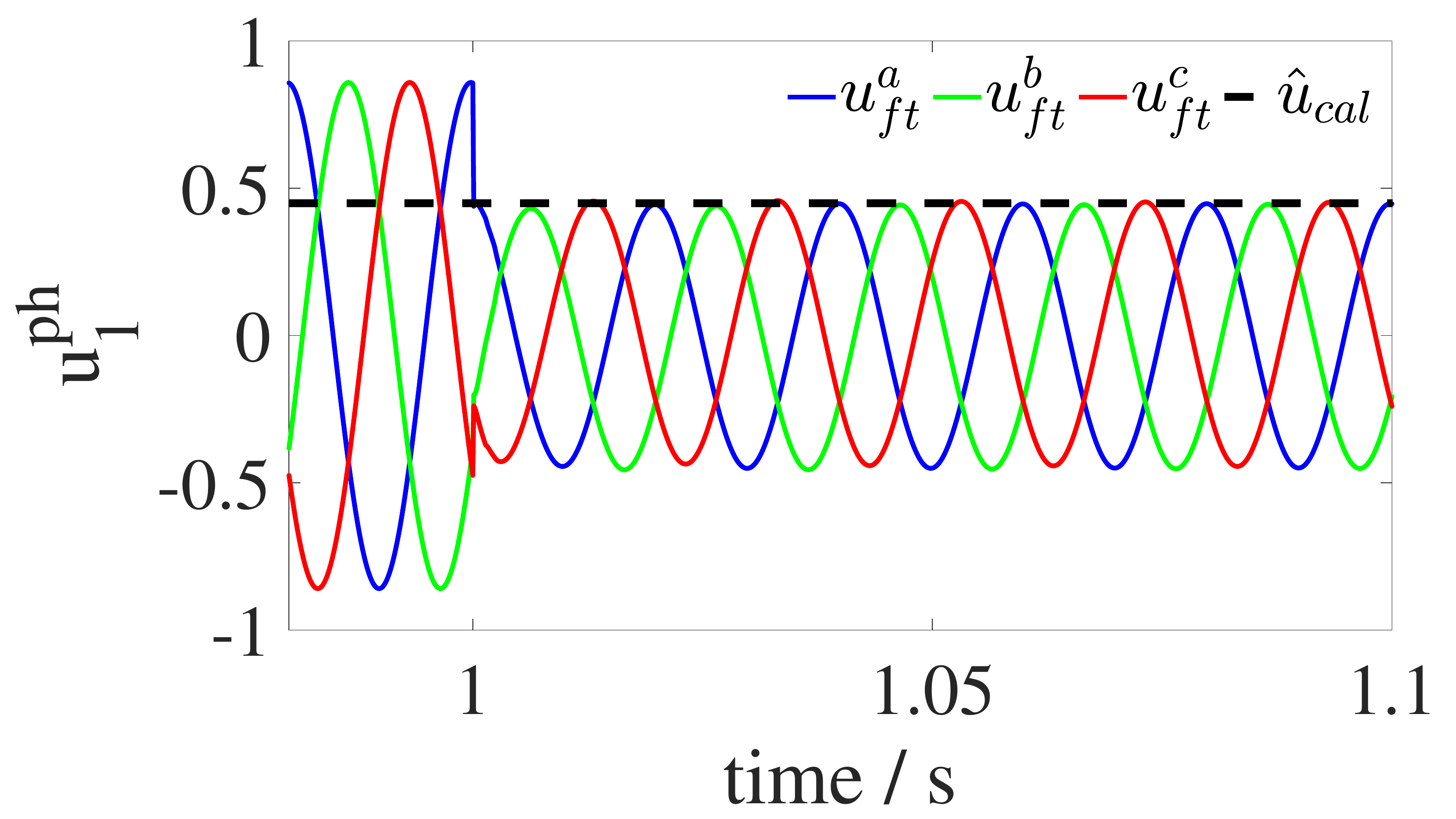}
	}
	\subfigure[VSC1 Power-3P2G.]{
		\includegraphics[width=0.225\textwidth]{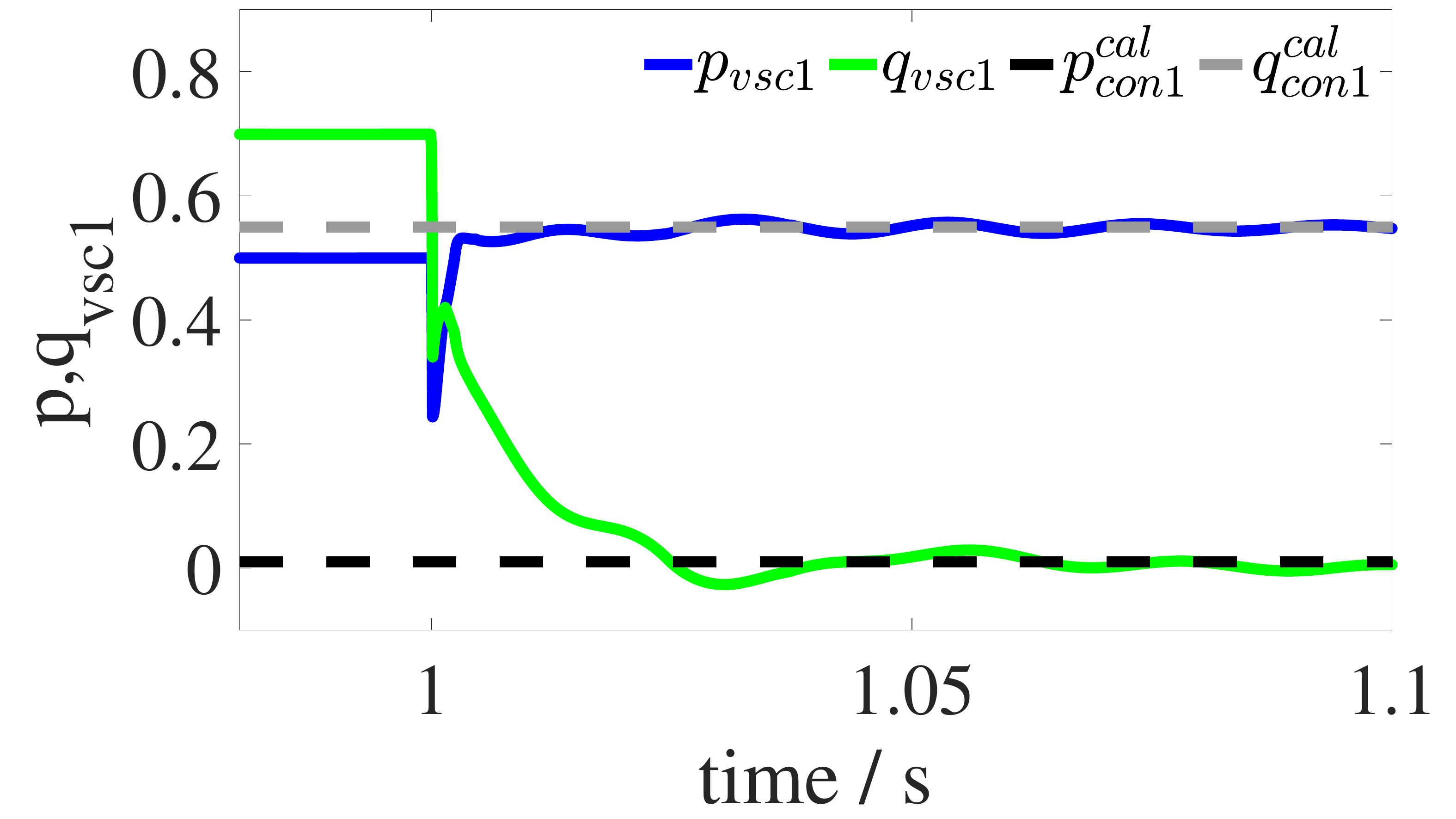}
	}
	\quad
	\subfigure[Fault voltage-P2P.]{
		\includegraphics[width=0.225\textwidth]{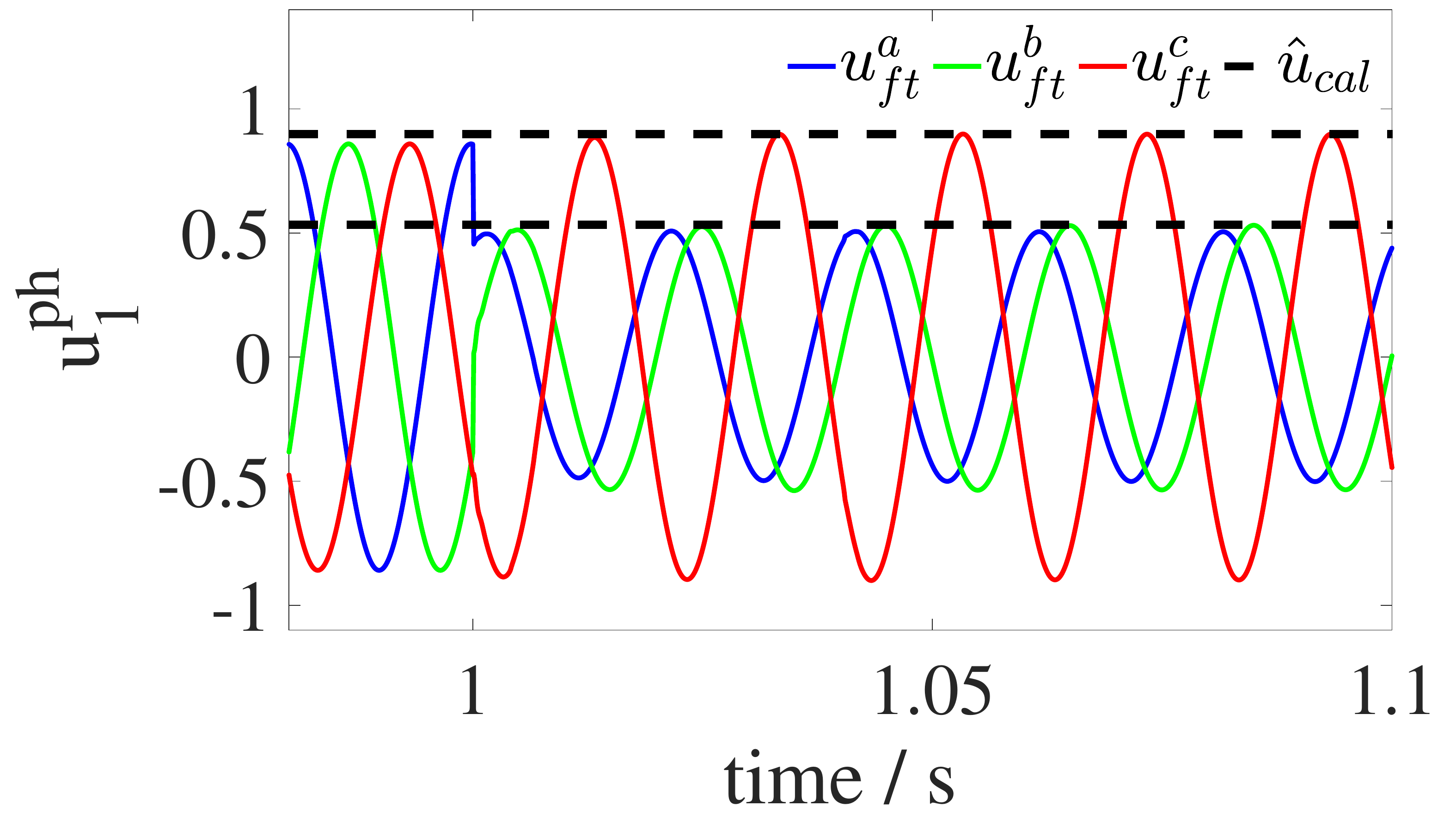}
	}
	\subfigure[VSC1 Power-P2P.]{
		\includegraphics[width=0.225\textwidth]{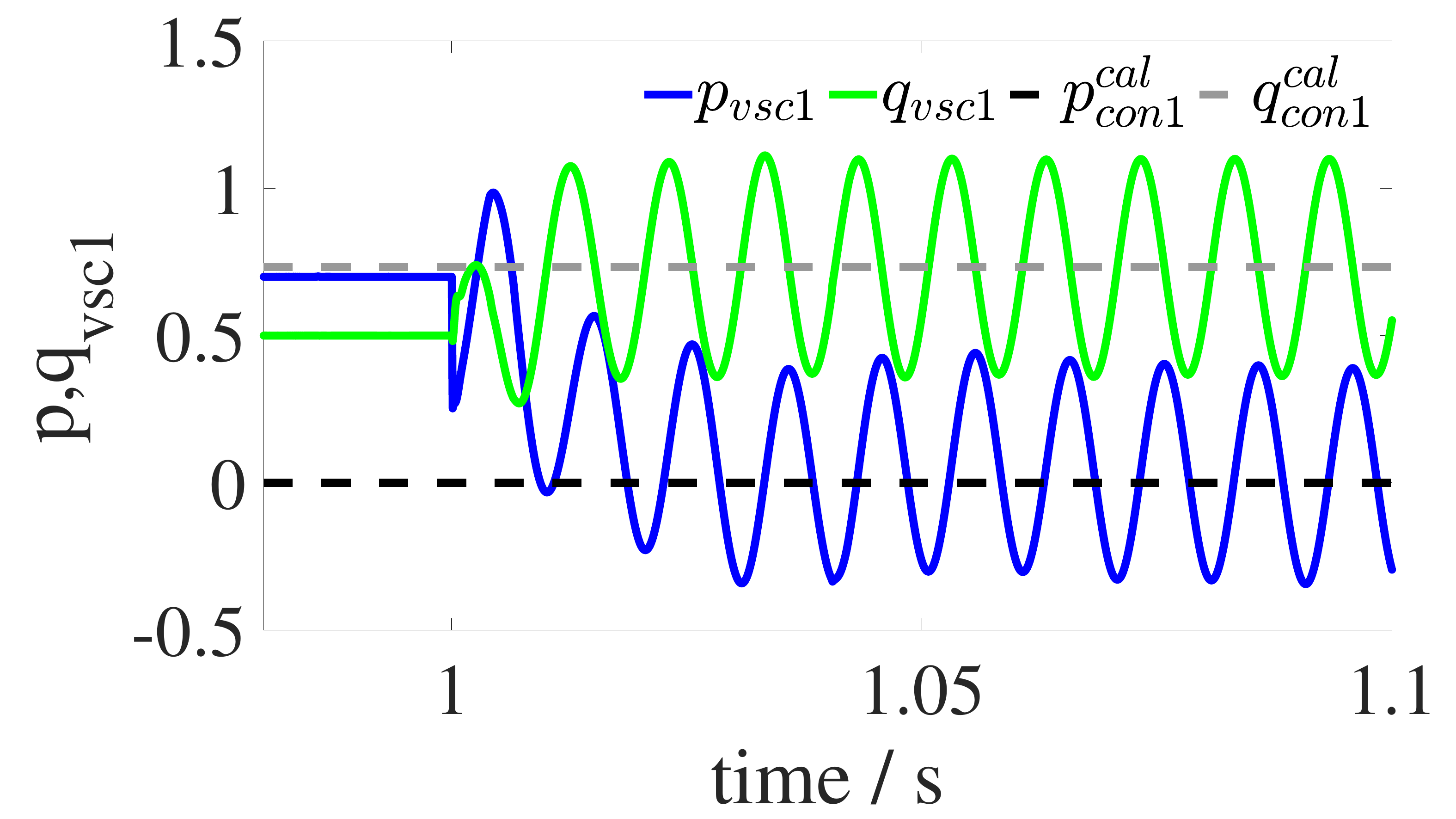}
	}
	\quad
	\subfigure[Fault voltage-1P2G.]{
		\includegraphics[width=0.225\textwidth]{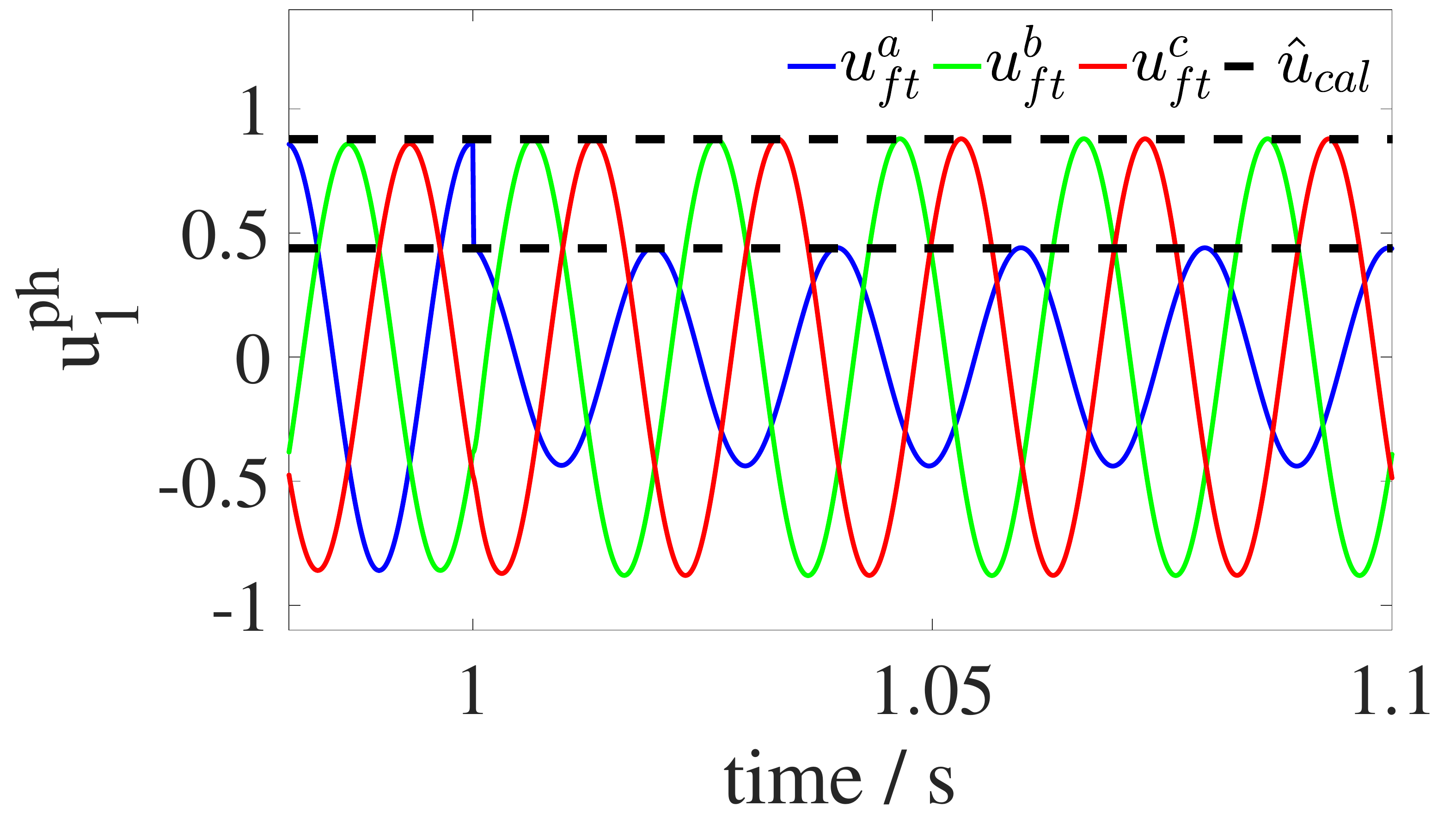}
	}
	\subfigure[VSC1 Power-1P2G.]{
		\includegraphics[width=0.225\textwidth]{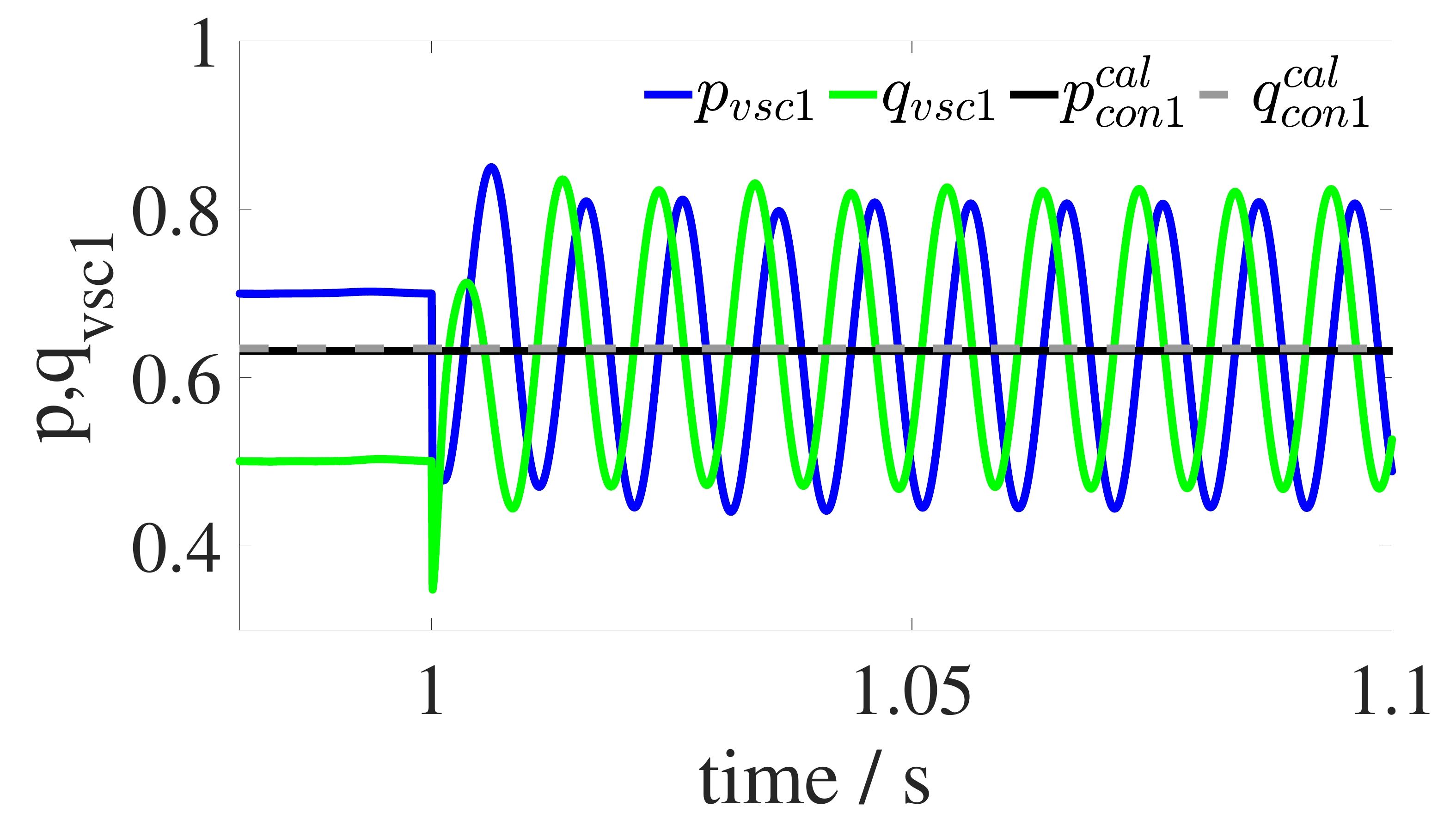}
	}
	\quad
	\vskip 0cm
	\caption{Dynamic Validation of Short-Circuit Calculation Results for Test System 1}\label{fig:Simulation Test 1}
\end{figure}

Dynamic simulations have been implemented in MATLAB Simulink in order to validate the obtained short-circuit equilibrium points. In particular, the fault impedance has been inserted to the fault location at the time of 1 s, which represents the fault event to Test System 1. The time-domain results are shown in Fig~\ref{fig:Simulation Test 1} in terms of the waveform of the fault voltage and the power injections from VSC1 for each tested type of fault. The steady-state calculated value in terms of the peak value of each phase voltage, $\hat{u}_{1}^{a,b,c}$, and the constant element of VSC1 power injection, $p_{con1}$ and $q_{con1}$, are also marked with dashed curves in Fig~\ref{fig:Simulation Test 1}. In can be observed that the obtained short-circuit equilibrium points are well-matched with dynamic simulation results, which proves the effectiveness of the proposed methodology. 

\subsection{Test System 2}

\begin{figure}[!htb]
	\centering
	\includegraphics[width=0.4\textwidth]{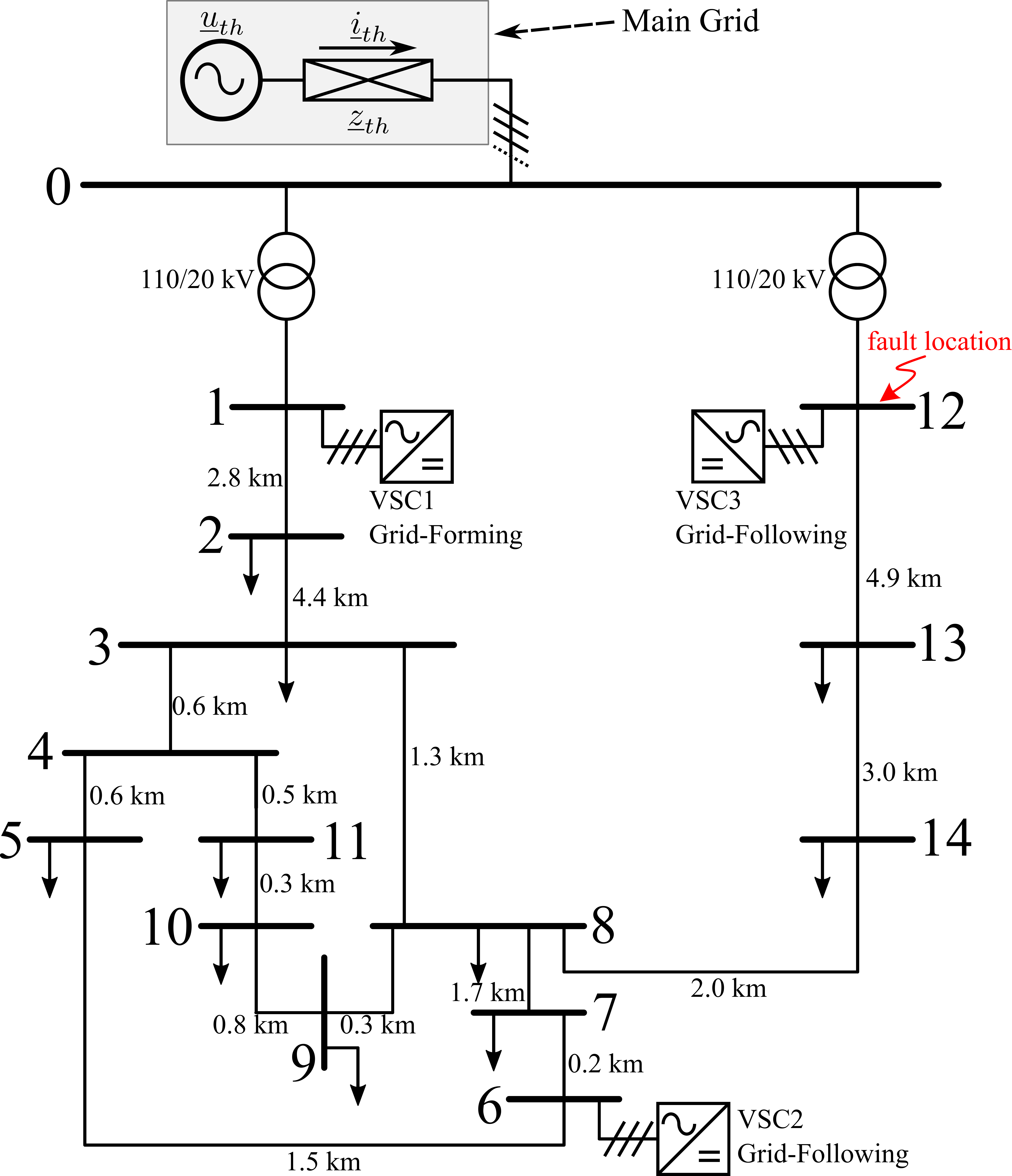}
	\caption{Scheme of Test System 2}
	\label{fig:Test2 Scheme}
\end{figure}

The second test system has been built based on the CIGRE benchmark model of European MV distribution system \cite{B575,9504581} with the scheme shown in Fig.~\ref{fig:Test2 Scheme}. The main AC grid, which is in four-wire configuration, is represented with the Thévenin equivalent and the frequency characteristic of the AC grid is included with the frequency-droop equations expressed as follows \cite{9504581}:
\begin{equation}
	\label{eq:ac grid frequency}
	\omega=\omega_0+k_{\omega-th} ({p_{con-th}-p_{0-th}})
\end{equation} 

\noindent where $k_{\omega-th}$ and $p_{0-th}$ are droop gain and dispatched active power for main grid frequency characteristic, $p_{con-th}$ is the constant element of active power injected from the main AC grid into the studied distribution system, which can be expressed as follows:
\begin{equation}
	\label{eq:ac grid active power}
	p_{con-th}={{u_{thx}^+ i_{thx}^+} + {u_{thy}^+ i_{thy}^+}} + {{u_{thx}^- i_{thx}^-} + {u_{thy}^- i_{thy}^-}}
\end{equation}

\FloatBarrier

Three VSCs with a three-wire connection are included in the studied distribution system, which are respective operating in PQ, PV and grid-forming modes. A frequency droop control has also been adopted by VSC1 (which is in grid-forming mode) as expressed in \eqref{eq:VSC_GF} to regulate the active power sharing between the converter and the main grid. The loads in the original distribution system are modeled with the equivalent impedance for short-circuit calculation, which is quantified under the nominal voltage. The circuit impedance is assumed as invariant for different grid frequency values \cite{Castro2012}. This approximation is acceptable for a low frequency deviation from the nominal value. 

Short-circuit has been tested at bus 12 (corresponding to the fault location indicated in Fig~\ref{fig:Test2 Scheme}) by inserting the fault impedance of $\underline{z}_{ft}=j0.05$ pu. In particular, three different types of fault have been tested. The identified results are shown in Table~\ref{table:Test2 Results} for each tested short-circuit fault. VSCs in the studied distribution system shows different current-saturation states with different fault types. In particular, all the three VSCs are fully saturated during the 3P2G fault, which leads to a low fault voltage in positive sequence. With P2P and 1P2G faults, VSC1, which is in grid-forming, remains unsaturated while the two grid-following converters are operating in different states. 

\begin{table}[!htb]
	\begin{center}
		\caption{Identified Short-Circuit Equilibrium Points of Test System 2}\label{table:Test2 Results}
		\scriptsize
		\begin{threeparttable}
			\begin{tabular}{m{1.6cm}<{\centering}|m{1.7cm}<{\centering}m{1.7cm}<{\centering}m{1.7cm}<{\centering}}
				\toprule
				Type of Fault & 3P2G & P2P & 1P2G   \\
				\midrule
				No. of Ite. $n_t$ & 2 & 2 & 3   \\
				\hline
				$t_{comp}$ ($s$) & 0.11 & 0.10 & 0.16   \\
				\hline
				\scalebox{0.95}{VSC1 State} & FSS & USS & USS   \\
				\hline
				\scalebox{0.95}{VSC2 State} & FSS & FSS & PSS   \\
				\hline
				\scalebox{0.95}{VSC3 State} & FSS & PSS & USS   \\
				\toprule
				$\omega$ & $0.995$ & $0.998$ & $0.996$  \\
				\hline
				$\underline{u}_{12}^+$ & $0.345\angle 6.9\si{\degree}$ & $0.753\angle -2.8\si{\degree}$ & $0.866\angle -3.9\si{\degree}$  \\
				\hline
				$\underline{u}_{12}^-$ & $0$ &\scalebox{0.95}{$0.594\angle -124.4\si{\degree}$} & $0.309\angle 174.
				0\si{\degree}$  \\
				\hline
				$\underline{u}_{12}^0$ & $0$ & $0$ & $0.309\angle 174.0\si{\degree}$  \\
				\hline
				$i_{vsc1}^+$ & $2.5$ & $1.547$ & $1.420$  \\
				\hline
				\scalebox{0.95}{$p_{con1}+jq_{con1}$} & $0.715+j1.935$ & $1.169+j1.014$ & $1.408+j0.183$  \\
				\hline
				$i_{vsc2}^+$ & $1$ & $1$ & $1$  \\
				\hline
				\scalebox{0.95}{$p_{con2}+jq_{con2}$} & $0+j0.661$ & $0+j0.935$ & \scalebox{0.95}{$-0.285+j0.927$}  \\
				\hline
				$i_{vsc3}^+$ & $1$ & $1$ & $0.957$  \\
				\hline
				\scalebox{0.95}{$p_{con3}+jq_{con3}$} & $0+j0.345$ & \scalebox{0.95}{$-0.664+j0.355$} & \scalebox{0.95}{$-0.800+j0.213$}  \\
				\bottomrule
			\end{tabular}
		\end{threeparttable}
	\end{center}
\end{table}

\begin{figure}[!htb]
	\centering
	\subfigure[Fault voltage-3P2G.]{
		\includegraphics[width=0.225\textwidth]{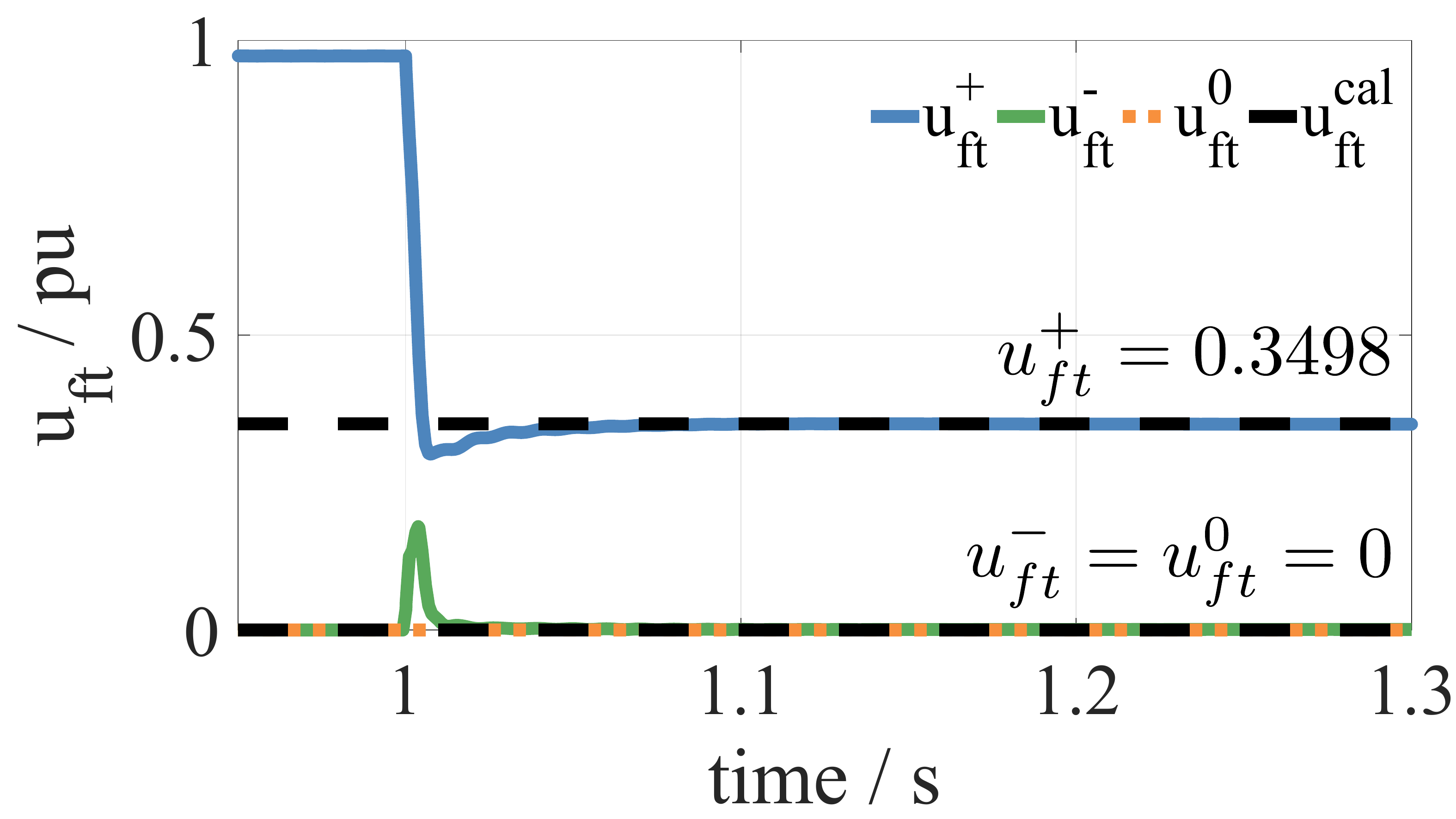}
	}
	\subfigure[Grid frequency-3P2G.]{
		\includegraphics[width=0.225\textwidth]{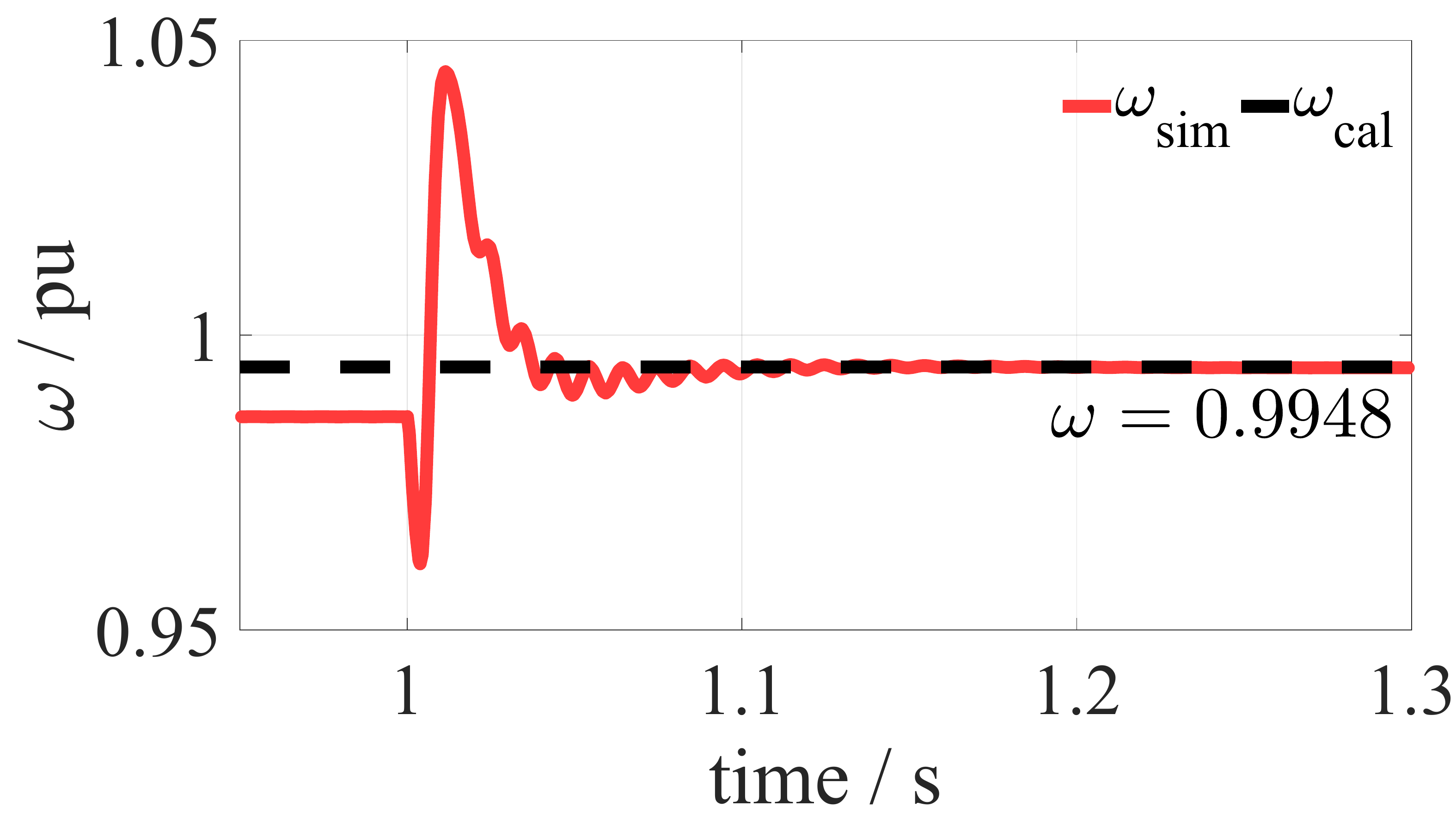}
	}
	\quad
	\subfigure[Fault voltage-P2P.]{
		\includegraphics[width=0.225\textwidth]{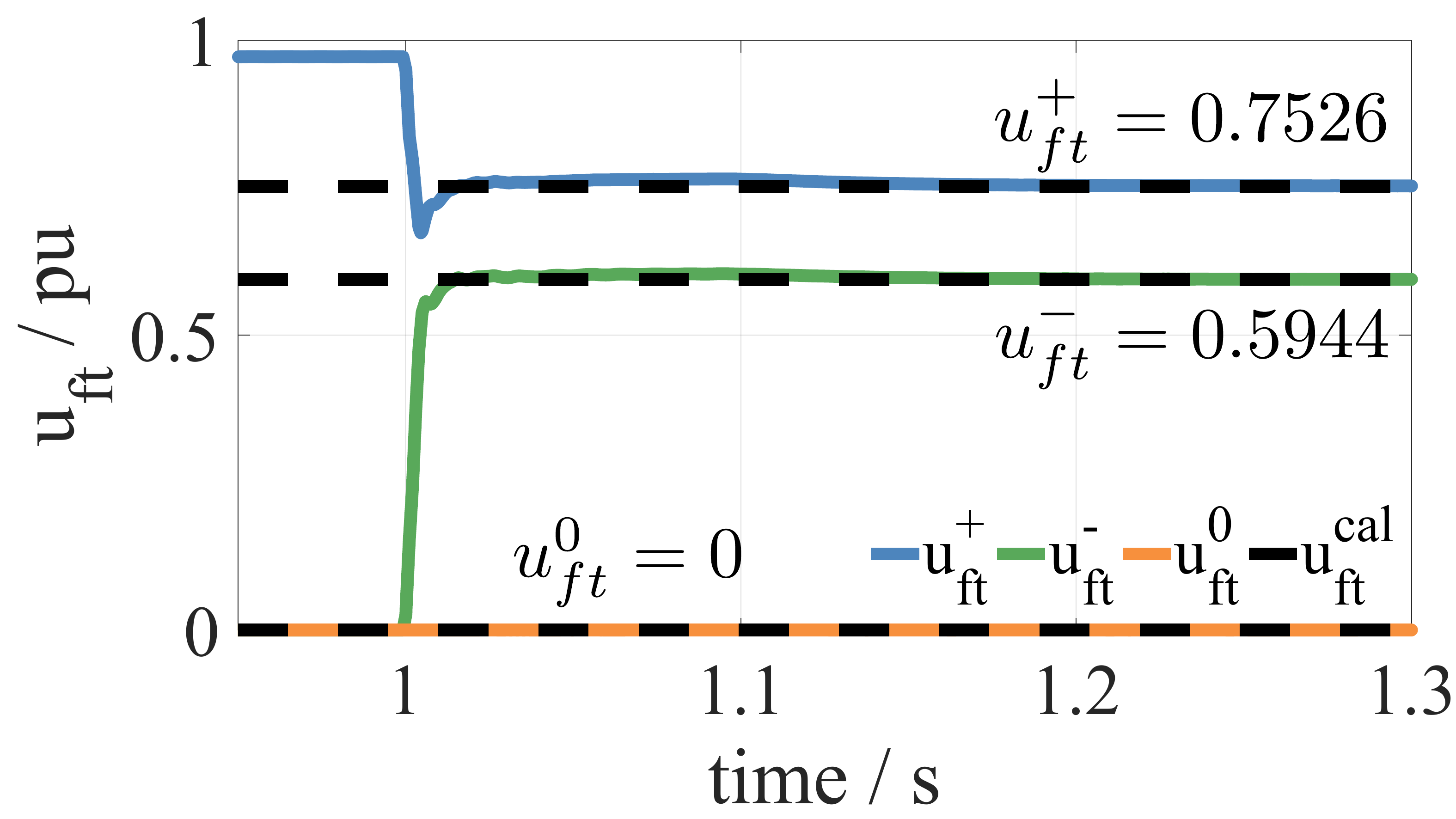}
	}
	\subfigure[Grid frequency-P2P.]{
		\includegraphics[width=0.225\textwidth]{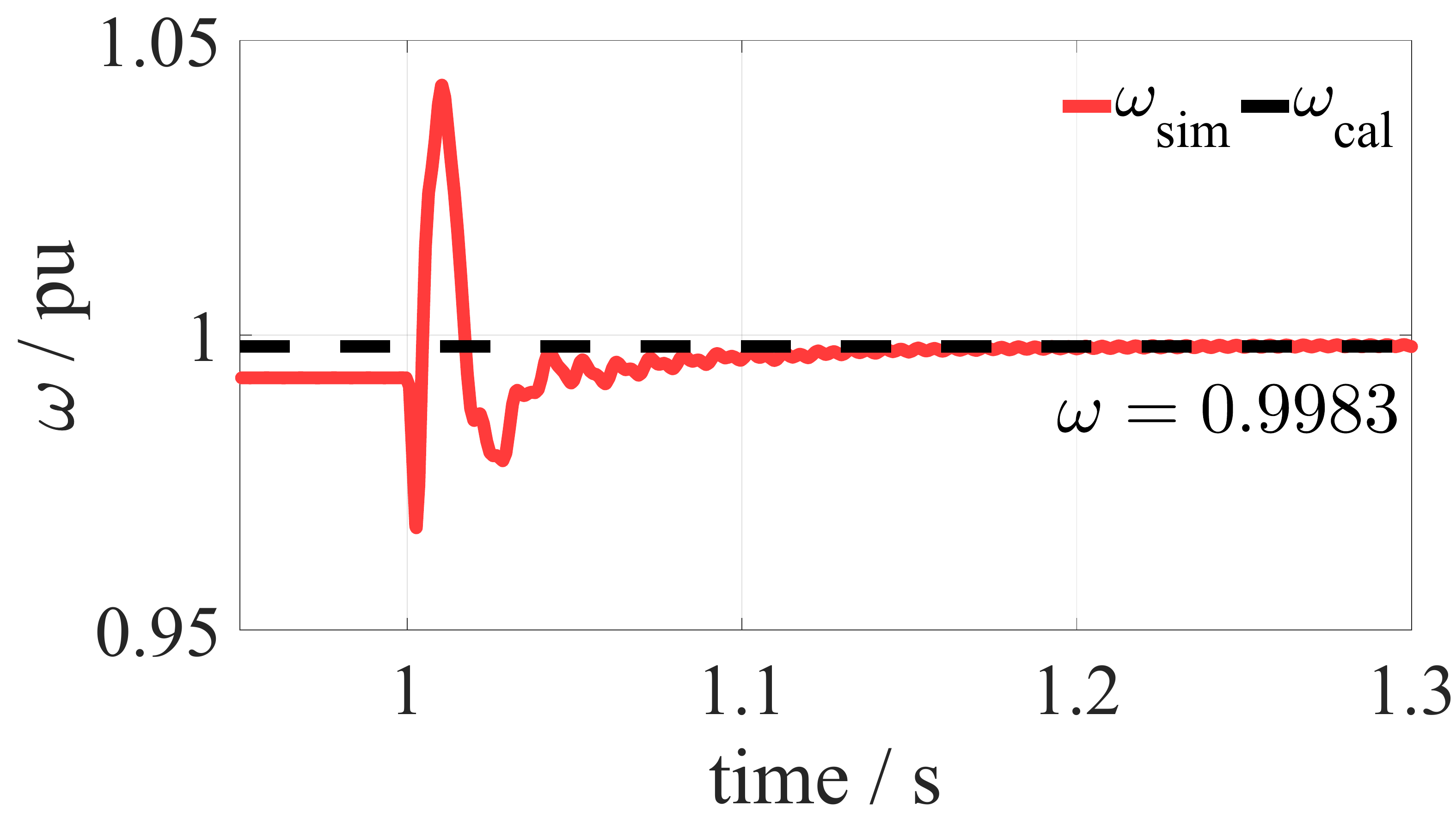}
	}
	\quad
		\subfigure[Fault voltage-1P2G.]{
		\includegraphics[width=0.225\textwidth]{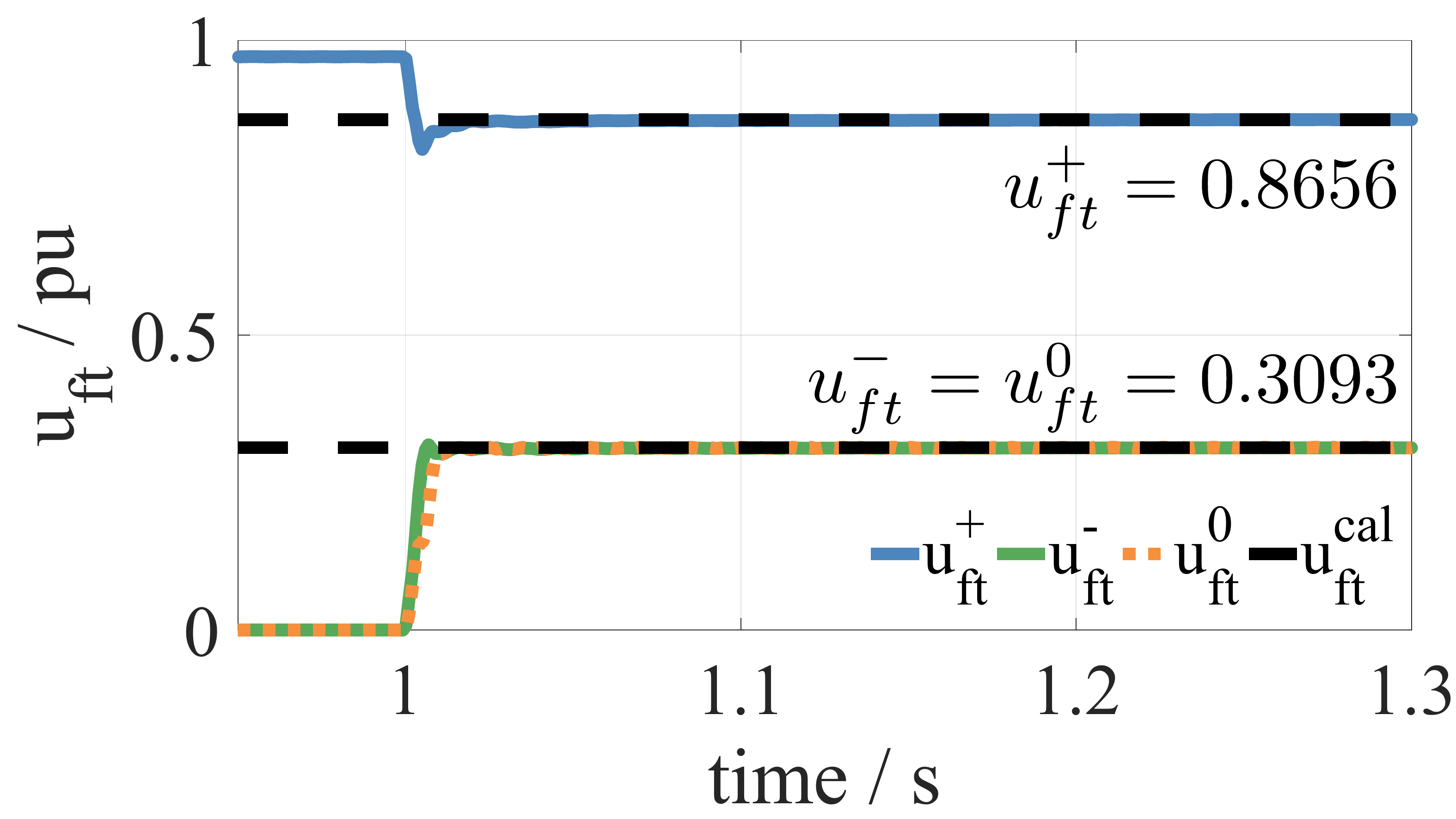}
	}
	\subfigure[Grid frequency-1P2G.]{
		\includegraphics[width=0.225\textwidth]{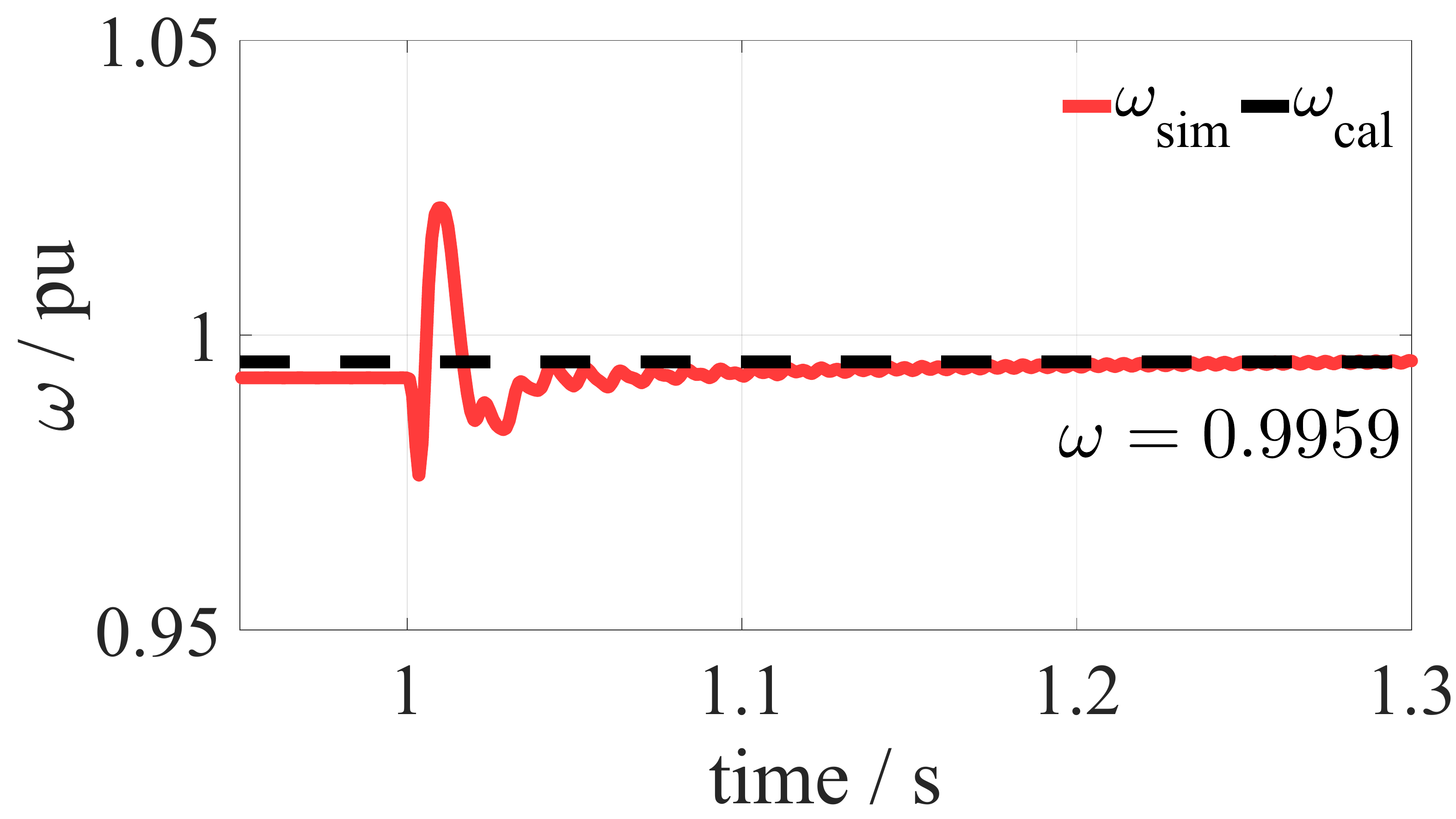}
	}
	\quad
	\vskip 0cm
	\caption{Dynamic Validation of Short-Circuit Calculation Results for Test System 2}\label{fig:Simulation Test 2}
\end{figure}

Dynamic simulations have been implemented to validate the obtained short-circuit equilibrium points. The time-domain results are shown in Fig~\ref{fig:Simulation Test 2} in terms of fault voltage magnitudes (sequence value) and the grid frequency for each fault scenario. The calculated steady-state values are marked with blacked dashed lines, which match with the dynamic simulation results. Also, the dynamic validation takes 2 minutes to perform simulations for each short-circuit equilibrium point to be analyzed. As a comparison, the proposed steady-state calculation can obtain the short-circuit equilibrium point in only 2 or 3 outer loop iteration with the computation time of less than 0.2 s, which shows significant advantage in efficiency compared to dynamic studies. 

\section{Conclusion}

This paper presents a different methodology for short-circuit calculation of power systems dominated by power electronics converters that is suitable for different types of fault. The 
equivalent model of power converters, which includes a specific control strategy an various possible current-saturation states, is included in the system formulation. An iterative algorithm has been proposed in order to efficiently identify the current-saturation states of all converters in the studied system for a specific fault scenario and therefore find the short-circuit equilibrium point that fulfills all converters' operation limits. 

Case studies in short-circuit calculation of test systems with VSCs proved the efficiency and effectiveness of the proposed methodology. A more comprehensive studies can be implemented on the short-circuit analysis with different options of converter control during the fault. In addition, further investigation can be carried out to develop the algorithm to solve the established steady-state formulation that could improve the computing efficiency. 

\section*{Appendix}

\subsection{Fortescue Transformation}

A three-phase phasor, $\underline{x}^{abc}$, can be represented with symmetrical sequence components applying the Fortescue transformation such that:
\begin{equation}
	\small
	\label{eq:Tn transformation}
	\left[ {\begin{array}{*{20}{c}}
			{{\underline{x}^ + }}\\
			{{\underline{x}^ - }}\\
			{{\underline{x}^0}}
	\end{array}} \right] = \underbrace{\frac{1}{3}\left[ {\begin{array}{*{20}{c}}
				1&\alpha &{{\alpha ^2}}\\
				1&{{\alpha ^2}}&\alpha \\
				1&1&1
		\end{array}} \right]}_{\mathbf{T^{-1}}}{\left[ {\begin{array}{*{20}{c}}
				{{\underline{x}^a}}\\
				{{\underline{x}^b}}\\
				{{\underline{x}^c}}
		\end{array}} \right]}
\end{equation}
\noindent where superscripts $+$, $-$ and $0$ respectively denote the positive, negative and zero sequence elements of a three-phase phasor $\underline{x}$. The symbol $\alpha$ is defined as: $\alpha=e^{-j2\pi/3}=-1/2-j\sqrt{3}/2$.  

\subsection{Test System Parameters}

The parameters of the two test systems analyzed in Section~\ref{sec:case study} are listed in Table~\ref{table:Test System parameters}. The circuit and load parameters of Test System 2 can be found in \cite{B575} and therefore are not included in this manuscript. 

\begin{table}[!htb]
	\scriptsize
	\begin{center}
		\caption{Parameters of Test Systems}
		\label{table:Test System parameters}
		\begin{tabular}{m{1cm}<{\centering}m{2.5cm}<{\centering}m{1cm}<{\centering}m{2.5cm}<{\centering}}
			\toprule
			Parameter & Value & Parameter & Value \\
			\midrule
			\multicolumn{4}{l}{\textbf{Test System 1}} \\
			\hline
			$u_{th}$ & $1$  & $\underline{z}_{th}$ & $0.01+j0.1$ \\
			\hline
			$p_{disp1}$ & $0.7$   & $q_{disp1}$ & $0.5$ \\
			\hline
			$i_{vsc1}^{\max}$ & $1$   & $k_{isp}$ & $2$ \\
			\midrule
			\multicolumn{4}{l}{\textbf{Test System 2}} \\
			\midrule
			$u_{th}$ & $1$  & $\underline{z}_{th}$ & $0.01+j0.10$\\
			\hline
			$\omega_{0}$ & $1$  & $k_{\omega -grid}$ & $0.05$\\
			\hline
			$p_{0-th}$ & $0$ & $u_{ref-gf}$ & $1$ \\ 
			\hline
			$k_{\omega}$ & $0.01$ & $p_{0}$ & $1$ \\
			\hline
			$i_{vsc1}^{\max}$ & $2.5$ & $p_{disp2}$ & $-0.794$\\
			\hline
			$u_{ref-PV}$ & $0.97$  & $i_{vsc2}^{\max}$ & $1$\\
			\hline
			$p_{disp3}$ & $-0.800$  & $q_{disp3}$ & $-0.023$\\
			\hline
			$i_{vsc3}^{\max}$ & $1$  & $k_{isp}$ & $2$\\
			\bottomrule
		\end{tabular}
	\end{center}
\end{table}

\bibliographystyle{ieeetr}
\bibliography{library}

\end{document}